\documentclass[aps,prl,twocolumn,groupedaddress,superscriptaddress,preprintnumbers,nofootinbib]{revtex4}

\usepackage[T1]{fontenc}
\usepackage{slashed}
\usepackage{times}

\usepackage{amssymb,amsfonts,amsmath,amsthm}
\usepackage{dsfont,bbm}
\usepackage{url}
\usepackage{graphicx}
\usepackage[colorlinks=true,linkcolor=blue,citecolor=blue]{hyperref}
\usepackage[usenames,dvipsnames]{color}

\newcommand{\nn}{\nonumber}

\usepackage[normalem]{ulem} 
\renewcommand\sout{\bgroup \color[rgb]{1,0.00,0.00} \ULdepth=-.5ex \ULset}

\newcommand{\eq}[1]{Eq.~\eqref{#1}}

\begin{document}

\preprint{NIKHEF 2015-36}

\title{Single spin asymmetries from a single Wilson loop}

\author{Dani\"el~Boer}
\affiliation{\normalsize\it  Van Swinderen Institute for Particle Physics and Gravity, University of Groningen, Nijenborgh 4, 9747 AG Groningen, The Netherlands}

\author{Miguel~G.~Echevarria}
\affiliation{\normalsize\it Nikhef and Department of Physics and Astronomy, VU Amsterdam,  De Boelelaan 1081, NL-1081 HV Amsterdam, The Netherlands}

\author{Piet~J.~Mulders}
\affiliation{\normalsize\it Nikhef and Department of Physics and Astronomy, VU Amsterdam,  De Boelelaan 1081, NL-1081 HV Amsterdam, The Netherlands}

\author{Jian~Zhou}
\affiliation{\normalsize\it School of physics, and Key Laboratory of
Particle Physics and Particle Irradiation (MOE), Shandong
University, Jinan, Shandong 250100, CHINA}
\affiliation{\normalsize\it
 Nikhef and Department of Physics and Astronomy, VU Amsterdam,  De Boelelaan 1081, NL-1081 HV Amsterdam, The Netherlands}

\begin{abstract}
We study the leading-power gluon transverse momentum dependent
distributions (TMDs) of relevance to the study of asymmetries in the scattering
off transversely polarized hadrons. Next-to-leading-order
perturbative calculations of these TMDs show that at large
transverse momentum they have common dynamical origins, but that in
the limit of small longitudinal momentum fraction $x$ only one
origin remains. We find that in this limit only the dipole-type gluon
TMDs survive and become identical to each other.
At small $x$ they are all given by the expectation value of a single
Wilson loop inside the transversely polarized hadron, the so-called
spin-dependent odderon. This universal origin of transverse spin
asymmetries at small $x$ is of importance to current and future
experimental studies, paving the way to a better understanding of
the role of gluons in the three-dimensional structure of
spin-polarized protons.
\end{abstract}

\maketitle

Scattering off protons in high energy particle collisions such as
performed at RHIC and LHC, can be described as scattering off quarks
and gluons inside the proton. As the energy of the scattering
process increases, the gluons play an increasingly important role,
as reflected by a fast growing gluon density inside the proton. The
limit of high gluon density is the one in which the gluons carry
only a very small fraction, $x$, of the longitudinal momentum of the
proton. It is expected on theoretical grounds that the gluon density
will not increase without bound towards small $x$, but rather that
it will saturate. Because of the dominance of gluons over quarks and
the expected gluon saturation, the description of scattering
processes considerably simplifies in the small-$x$ limit. A
complication that remains in this limit though, is that gluonic
effects do not manifest themselves in the same way in all processes.

It has recently become clear that the transverse momentum distribution of
gluons inside the proton is not a unique quantity. Different
processes may probe different distributions and
thus yield different answers. This has become apparent from studies
of the unpolarized gluon distribution in the small-$x$ regime
\cite{Kharzeev:2004yx,Dominguez:2010xd,Dominguez:2011wm} and,
independently, from studies of spin effects in high-energy
scattering processes. Scattering experiments involving a
spin-polarized proton exhibit large asymmetries in the production of
final state particles
\cite{Antille:1980th,Adams:1991cs,Adams:1991rw,Krueger:1998hz,Adams:2003fx,Adler:2005in,Arsene:2008aa,Abelev:2008af,Adamczyk:2012xd}.
Theoretical studies of these still largely ununderstood single
spin asymmetries (SSA) led to the insight that TMDs of both quarks and
gluons are sensitive to the flow of the color charge of quarks and
gluons in a process and hence that they are in general process
specific, i.e.\ nonuniversal~\cite{Collins:2002kn}. TMD studies of
the color flow dependence are generally not performed in the high
gluon density region. In~\cite{Dominguez:2010xd,Dominguez:2011wm},
the two types of treatments were connected for the case in which
neither the proton nor the gluons are spin-polarized.
For transversely polarized protons the connection between the TMD and
small-$x$ formalisms has so far not been made. This is our aim here.

We will study the (T-odd) gluon TMDs inside a proton that is polarized transversely to its momentum direction and consider the limit of high gluon density or small $x$ fraction.
We then connect the results to those that arise in a small-$x$ treatment and observe that the two pictures are fully compatible, despite the initial mismatch in the number of distributions.
Surprisingly, unlike the unpolarized case, we find that there is in fact only one type of gluon correlation to consider in the transversely polarized case in the limit of small $x$, thereby reducing the high degree of nonuniversality~\cite{Buffing:2013kca} to a single, universal distribution.
The distribution is linked to what has been discussed in the literature under the name of spin-dependent odderon.
Although gluon induced SSA are likely smaller than valence quark ones,
the universality of gluon effects in the transverse spin case is of high experimental interest, as it can be investigated
 at RHIC using collisions of polarized protons on heavy ions and possibly directly compared to data from proposed experiments at an Electron-Ion Collider (EIC) or a polarized fixed-target experiment at LHC called AFTER@LHC.

The interplay of spin/TMD physics and small-$x$ physics is a topical
issue. Recent developments in this direction include modeling
nuclear quark TMDs using quasi-classical
methods~\cite{Schafer:2013mza,Kovchegov:2013cva,Kovchegov:2015zha},
the study of linear gluon polarization in the
small-$x$ formalism~\cite{Metz:2011wb,Dominguez:2011br} and the
evolution of gluon distributions from moderate to low $x$~\cite{Mueller:2013wwa,Schafer:2013opa,Kovchegov:2015zha,Balitsky:2015qba}. Phenomenological studies of T-odd gluon TMDs have been performed in
\cite{Qiu:2011ai,Anselmino:2015eoa,D'Alesio:2015uta,Ma:2015vpt} without including process dependence.
Spin asymmetries in $pA$ collisions have been
investigated extensively~\cite{Boer:2006rj,Kang:2011ni,Kovchegov:2012ga,Kang:2012vm,Akcakaya:2012si,Altinoluk:2014oxa,Kotko:2015ura,Schafer:2014zea}.
The longitudinal proton spin distribution $\Delta g$ or $g_{1L}$ has been studied in the
small-$x$ regime in~\cite{Bartels:1995iu,Maul:2001uz,Kovchegov:2015pbl}.
The transverse spin case turns out to be quite different.

This Letter is structured as follows. First, we present the
calculation of the large transverse momentum tail of the gluon TMDs
of relevance for single transverse spin asymmetries and consider the
small-$x$ limit. A reduction from three independent TMDs to just one
is observed. Subsequently we connect this distribution to the
spin-dependent odderon distribution arising in small-$x$ studies,
 finding full consistency among the results. We end with a discussion
and conclusions.

The information on gluon TMDs inside a transversely polarized hadron
(with spin vector $S_T$) is formally encoded in the following matrix
element~\footnote{For the sake of simplicity we omit a soft factor
in the properly defined gluon
TMDs~\cite{GarciaEchevarria:2011rb,Echevarria:2015uaa,Collins:2011zzd},
as this does not affect the results of this work.},
\begin{eqnarray}
&&\Gamma^{\mu \nu [U,U']} =
\frac{1}{xP^+}\int \frac{dy^- d^2y_T}{(2\pi)^3} e^{ik \cdot y}
\nonumber \\
&\times& \langle P, S_T | 2 {\rm Tr} \left [ F_{T}^{+\nu}(0) U
F_{T}^{+\mu}(y)U' \right ] |P, S_T \rangle \big|_{y^+=0} ,
\label{gmat}
\end{eqnarray}
where $U$ and $U'$ are process dependent gauge links in the
fundamental representation. At leading power, this correlator can be
parameterized by six independent tensor
structures~\cite{Mulders:2000sh},
\begin{multline}
\Gamma^{\mu \nu}= \delta_T^{\mu \nu} f_1^g- \left
(\frac{2k_T^\mu k_T^\nu}{k_T^2}+\delta_T^{\mu\nu} \right
)  h_1^{\perp g}
\\
- \delta_T^{\mu \nu} \frac{\epsilon_{T\alpha \beta} k_T^\alpha S_T^\beta}{M}
f_{1T}^{\perp g}
-i \epsilon^{\mu \nu}_T \frac{k_T \cdot S_T}{M} g_{1T}^g
 \\
-\frac{ \tilde k_T^{ \{\mu } S_T^{\nu \} } +\tilde S_T^{ \{\mu }
k_T^{\nu \} }}{2M}h_{1T}^{ g} + \frac{ \tilde k_T^{ \{\mu }
k_T^{\nu \} } }{ k_T^2} \frac{k_T \! \cdot \!\! S_T}{ M}
h_{1T}^{\perp g},
\end{multline}
where all six TMDs are functions of $x$ and $k_T^2$,
$\epsilon_T^{\mu \nu}=\epsilon^{\rho \sigma \mu \nu} n_\rho
p_\sigma$,
 with $\epsilon_T^{12}=1$  and $\delta_T^{\mu \nu}=-g^{\mu\nu}+p^{ \{\mu} n^{\mu \} } /p \cdot n$.
We also used short hand notations like
$\tilde k_T^\nu=\epsilon_T^{\mu \nu} k_{T\mu}$. Note that the
normalizations for gluon TMDs $h_1^{\perp g}$, $h_{1T}^{\perp
g}$ and $h_{1T}^{ g}$ are slightly different from the ones used in~\cite{Mulders:2000sh,Meissner:2007rx}, because our results
suggest it to be a more natural choice.

The first two gluon TMDs, $f_1^g$ and $h_1^{\perp g}$, are the
unpolarized and linearly polarized gluon distribution, respectively.
Among the four transverse spin dependent gluon TMDs, the three T-odd
gluon TMDs, $f_{1T}^{\perp g}$, $h_{1T}^{\perp g}$ and $h_{1T}^{g}$,
are relevant for the single spin asymmetry studies. As mentioned,
none of these TMDs is universal and should have a $[U,U']$ label, as
in general, different processes probe matrix elements with different
gauge links. Here we will restrict ourselves to the two most
important cases, involving only single future or past pointing
staple-like gauge links, denoted by $+$ and $-$, respectively.
In the notation of \cite{Bomhof:2007xt} there are two T-odd combinations labeled with $(f)$ and $(d)$,
 $\Gamma_{(f)}^{(T-{\rm odd})} =
(\Gamma^{[+,+\dagger]}-\Gamma^{[-,-\dagger]})/2$ and
$\Gamma_{(d)}^{(T-{\rm odd})} =
(\Gamma^{[+,-\dagger]}-\Gamma^{[-,+\dagger]})/2$.
For the unpolarized gluon distribution at small $x$ the first type
is usually referred to as the Weizs\"{a}cker-Williams (WW)
distribution, while the latter one is commonly known as the dipole
gluon distribution~\cite{Dominguez:2010xd,Dominguez:2011wm}. Here we
will also refer to TMDs with the superscript "$(f)$" as WW type
distributions, and with a "$(d)$" as dipole type gluon TMDs.
The transverse moments of the $(f)$ and
$(d)$ type functions are related to single gluon pole matrix
elements with different color structures, $f^{abc}$ and $d^{abc}$,
respectively.

T-odd TMDs with more complicated link structures can in principle arise,
but do not in any currently known TMD-factorizing process. At small $x$ some processes
can become effectively TMD factorizing, where additional distributions could
enter \cite{Dominguez:2010xd}. These differ from the $(f)$ and $(d)$
type distributions by terms of subleading order in $1/N_c$, which can be calculated
within a small-$x$ formalism \cite{Akcakaya:2012si,Kotko:2015ura}. So if relevant at all, they can to some extent
be related to the distributions considered here (also following the methods of \cite{Buffing:2013kca}).

In analogy to the T-odd quark TMDs~\cite{Ji:2006ub,Zhou:2008fb,Dai:2014ala},
all three T-odd gluon TMDs can
be perturbatively calculated in the collinear twist-3 formalism at
large transverse momentum. The hard coefficients entering in these
expressions are usually different for different gauge links
appearing in the gluon matrix element given in Eq.~(\ref{gmat}).
Here we will present the results for the $(f)$ and $(d)$ type
functions that are $C$-even and $C$-odd, respectively. The T-odd
collinear twist-3 functions that appear in the large $k_T$
tail expressions are (chiral-even) quark-gluon and tri-gluon
Qiu-Sterman functions~\cite{Qiu:1991pp,Ji:1992eu,Braun:2009mi}, with
matching $C$-parity.
Chiral-odd quark gluon Qiu-Sterman functions are suppressed for gluon TMDs.

\begin{figure}[t]
\begin{center}
\includegraphics[width=7cm]{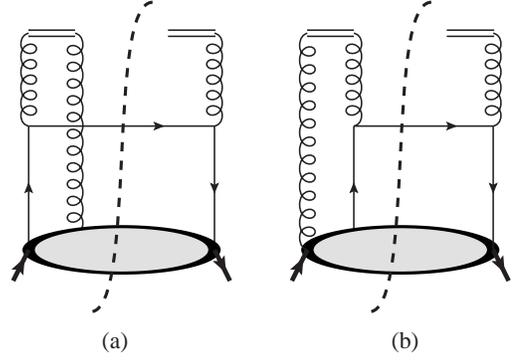}
\\
(a)\hspace{3.5cm}(b)
\end{center}
\caption{\it
Diagrams contributing to T-odd gluon TMDs at large transverse momentum in the flavor-singlet case.
(a): soft-gluon pole contribution. (b): hard-gluon pole contribution.
(Mirror diagrams are not shown).
}
\label{fig}
\end{figure}

The perturbative calculation follows a similar procedure as
in~\cite{Ji:2006ub}, cf.\ Fig.\ \ref{fig}. The WW type gluon Sivers function
$f_{1T}^\perp$ has been computed in the quark channel in terms of
the quark-gluon Qiu-Sterman function $T_{F,q}$ in~\cite{Schafer:2013wca},
\begin{eqnarray}
&& \!\!\!\!\!\!\!\! f_{1T}^{\perp g/q\,(f)} (x,k_T^2)= C_1
\frac{M}{k_T^4} \int_{x}^{1}  \frac{dz}{z} \sum_{q+\bar{q}}
\nonumber \\
&& \times
\left \{T_{F,q} (z,z) \frac { 1+(1-\xi)^2}{\xi} -T_{F,q}(z,z-x)  \frac{2-\xi}{\xi}  \right \},
\label{fsivers}
\end{eqnarray}
where $\xi=x/z$ and $C_1=\frac{N_c}{2} \frac{\alpha_s}{2\pi^2}$. The
notation $T_F$ is the same as in~\cite{Ji:2006ub}. The
$\sum_{q+\bar{q}}$ indicates that the sum runs over all quark
flavors and anti-flavors. Here, a factor $-g$ is included in the
definition of the anti-quark Qiu-Sterman function, such that it
satisfies $T_{F,\bar
q}(x_1,x_2)=T_{F,q}(-x_1,-x_2)$~\cite{Zhou:2015lxa}. The soft gluon
pole contribution (the first term within the brackets) is generated
by the diagram shown in Fig.~\ref{fig}(a), and the hard gluon pole
contribution (the second term) arises from Fig.~\ref{fig}(b). This expression
can be related to the one in \cite{Yuan:2008vn}.
Throughout this paper we will neglect the contribution from the
antisymmetric partner of the Qiu-Sterman function, $\tilde
T_F$~\cite{Braun:2009mi}, which becomes suppressed in the small-$x$
regime, regardless of the gauge link structure, as it is
antisymmetric in its two arguments (assuming it has no pole).

The gluon TMDs $h_{1T}^g$ and $h_{1T}^{\perp g}$ can be calculated
similarly at large transverse momentum. They turn out to possess the
same perturbative tail $1/k_T^4$ behavior as the Sivers
function, only differing in the hard coefficients:
\begin{eqnarray}\label{eq:h1h2f}
&& \!\!\!\!\!\!\!\! h_{1T}^{g/q\,(f)} (x,k_T^2) = C_1
\frac{M}{k_T^4} \int_{x}^{1}  \frac{dz}{z} \sum_{q+\bar{q}}
\nonumber \\ && \times \left \{T_{F,q} (z,z) \frac { 2-2\xi }{\xi}
-T_{F,q}(z,z-x) \frac{2-\xi}{\xi}  \right \},
\\
&& \!\!\!\!\!\!\!\! h_{1T}^{\perp g/q\,(f)} (x,k_T^2) =
 C_1  \frac{M}{k_T^4} \int_{x}^{1} \frac{dz}{z} \sum_{q+\bar{q}}
T_{F,q} (z,z) \frac {4-4\xi}{\xi} .
\end{eqnarray}
We note that the hard gluon pole contribution to $h_{1T}^{\perp g}$
is absent.

We now extrapolate these results to the small-$x$ limit.
For the WW type distributions, it is easy to see that in the small-$x$ limit the gluon
TMDs $f_{1T}^{\perp g/q\,(f)}$ and  $h_{1T}^{g/q\,(f)}$ vanish up to leading
logarithm $\ln\tfrac{1}{x}$ accuracy, due to the cancelation among
the soft-gluon and hard-gluon pole contributions.
The same cancelation occurs at small $x$ for the tri-gluon correlation
contribution: $f_{1T}^{\perp g/g\,(f)}\approx h_{1T}^{g/g\,(f)}\approx 0$.

The case of $h_{1T}^{\perp g(f)}$ is different, however. Combining
the small-$x$ limit of the quark channel in \eq{eq:h1h2f} with the
contribution of the gluon channel, it takes the form
\begin{eqnarray}
&&h_{1T}^{\perp g\, (f)} (x,k_T^2) \approx C_1  \frac{M}{k_T^4}
\frac{4}{x}
\nn\\
&& \times \int_{x\to 0}^{1}  dz \, \left \{ \sum_{q+\bar{q}} T_{F,q}
(z,z) + T_{G}^{(+)}(z,z)  \right \}
 \,,
\end{eqnarray}
where $T_{G}^{(+)}$ is the $C$-even tri-gluon
correlation~\cite{Ji:1992eu,Kang:2008qh,Beppu:2010qn}. This
particular integral vanishes as a consequence of transverse momentum
conservation, as it can be related (at tree level certainly
\cite{Boer:2003cm} and the relation is stable under QCD corrections
\cite{Zhou:2015lxa}) to the Burkardt sum rule for the first
transverse momentum of the Sivers TMD~\cite{Burkardt:2003yg}.
Therefore, for the $ h_{1T}^{\perp g(f)}$ case, the leading
logarithm contributions cancel out between the quark and gluon
channels.

Now we consider the dipole case. Again all three TMDs can be
dynamically generated by the Qiu-Sterman function, and possess the
same perturbative tail $1/k_T^4$. The result for the gluon
Sivers function in the quark channel is
\begin{eqnarray}
&& \!\!\!\!\!\!\!\! f_{1T}^{\perp g/q\,(d)} (x,k_T^2)=C_2
\frac{M}{k_T^4} \int_{x}^{1}  \frac{dz}{z} \sum_{q-\bar{q}}
\nonumber
\\ && \times \left \{T_{F,q} (z,z) \frac { 1+(1-\xi)^2}{\xi}
+T_{F,q}(z,z-x)  \frac{2-\xi}{\xi}  \right \}, \label{dsivers}
\end{eqnarray}
where $C_2=\frac{N_c^2-4}{2N_c}  \frac{\alpha_s}{2\pi^2}$. The
$\sum_{q-\bar{q}}$ indicates that in this $C$-odd case the sum runs
over all quark flavors minus anti-flavors. Similarly, for the other
two gluon TMDs we find
\begin{eqnarray}
&&h_{1T}^{g/q\,(d)} (x,k_T^2) = C_2  \frac{M}{k_T^4}
\int_{x}^{1} \frac{dz}{z} \sum_{q-\bar{q}}
\nn \\
&& \quad\quad\times
\left \{T_{F,q} (z,z) \frac { 2-2\xi}{\xi} +T_{F,q}(z,z-x)  \frac{2-\xi}{\xi}  \right \}
,
\nn\\
&& h_{1T}^{\perp g/q\,(d)} (x,k_T^2) = C_2 \frac{M}{k_T^4}
\int_{x}^{1} \frac{dz}{z} \sum_{q-\bar{q}}
\nn\\
&& \quad\quad\times
T_{F,q} (z,z)
\frac {4-4\xi}{\xi}
\,.
\end{eqnarray}
It is worth noting that as compared to the WW type distributions,
the overall color factor is different and the sign of the hard-gluon
pole contributions is reversed. The complete expressions for the
gluon channel ($g/g$) will be presented elsewhere. Here we only
present the extrapolation to the small-$x$ limit. In this limit all
three dipole type T-odd gluon TMDs take the same form in both the
quark and the gluon channel:
\begin{eqnarray}
& & f_{1T}^{\perp g\,  (d)}  \approx  h_{1T}^{g \, (d)} \approx h_{1T}^{\perp g \, (d)}
\approx \frac{M}{k_T^4} \frac{4}{x}
\nn\\
& & \times  \int_{x\to 0}^{1}  dz\, \left \{ C_2 \sum_{q-\bar{q}} T_{F,q} (z,z) +
C_1 T_{G}^{(-)} (z,z) \right \} \,,
\end{eqnarray}
where $T_{G}^{(-)}$ is the $C$-odd tri-gluon correlation defined in~\cite{Ji:1992eu,Kang:2008qh,Beppu:2010qn}. The splitting
kernel that appears in the above formula is identical to that for
the ordinary unpolarized gluon distribution for the gluon to gluon
channel. Therefore,  from these large $k_T$ expressions, we conclude
that the dipole type T-odd gluon TMDs are not necessarily suppressed
at small $x$ w.r.t.\ the unpolarized gluon distribution (which grows
very rapidly towards small $x$), but the WW type T-odd gluon TMDs
are. The additional $1/k_T^2$ suppression compared to the $1/k_T^2$
large-$k_T$ tail of the unpolarized gluon distribution does not mean
the TMDs are power suppressed at smaller $k_T$ too.

The existence of leading logarithm contributions in the dipole case
actually justifies a small-$x$ treatment of T-odd gluon
TMDs {\it without} the restriction of large $k_T$.
Our starting point is the T-odd part of the {\it dipole} type
gluon TMD matrix element (we will suppress the label $(d)$ from now
on):
\begin{eqnarray}
&& \!\!\!\!\!\!\!\!\!\!\!\!\!\! \Gamma^{\mu \nu}_{\rm T-odd} (x,
k_T; S_T)= \frac{1}{xP^+}\int \frac{dy^- d^2y_{T} }{(2\pi)^3} e^{ik
\cdot y} \nonumber \\ &&
 \!\!\!\!\!\!\! \!\!\!\! \times \left \{
\langle P, S_T |  {\rm Tr} \left [ F_{T}^{+ \nu}(0) U^{[+]}
 F_{T}^{+\mu}(y) U^{[-] \dag} \right .\ \right .\
\nonumber \\
&& \left .\  \left .
 \!\!\!\!\! \!\!\!\! \!\!\!\!
- F_{T}^{+\nu}(0) U^{[-]} F_{T}^{+\mu}(y)U^{[+]\dag} \right ]
 |P, S_T \rangle \right \}\big|_{y^+=0}.
\label{matrix1}
\end{eqnarray}
Next we approximate the exponential $e^{ik^+y^-}$ by $1$.
In~\cite{Dominguez:2010xd,Dominguez:2011wm} this is argued to be a
good approximation as long as $k^+=xP^+$ is very small, although
corrections may affect the rapidity evolution, as recently discussed
in~\cite{Balitsky:2015qba}. After making this approximation,
Eq.~(\ref{matrix1}) can be re-organized as
\begin{align}
&\!\! \Gamma^{\mu \nu}_{\rm T-odd}(x, k_T; S_T) = \frac{ k_T^\mu
k_T^\nu}{g^2 V xP^+}\int \frac{d^2y_{T}}{(2\pi)^3} e^{ik_T \cdot
y_T}
\nonumber \\
& \!\! \times
\langle P, S_T |  {\rm Tr} \left [
U^{[\Box]}(0_T,y_T) - U^{[\Box] \dag}(0_T,y_T) \right ]
 |P, S_T \rangle,
\label{matrix2}
\end{align}
where $V=\int dy^-$ and $U^{[\Box]}$ represents a rectangular Wilson loop with lightlike Wilson lines at transverse separation $y_T$.
To arrive at this equation we used translational invariance and
\begin{eqnarray}
&& \!\!\!\!\!\!\!
\partial_T^\mu U(y_T)=-ig\int_{-\infty}^{+\infty} d y^-
\nonumber \\ && \times U[-\infty^-,y^-;y_T] \partial_T^\mu
A_+(y^-,y_T) U[y^-,\infty^-;y_T],
 \end{eqnarray}
where $\partial_T^\mu A_+(y^-,y_T)$ is part of the gluon field
strength operator $F_{T}^{+\mu}$. The $\partial_+ A_T^\mu(y^-,y_T)$
part corresponds to the transverse gauge link at lightcone infinity
which can be neglected in a covariant gauge
calculation~\cite{Collins:2002kn}. The remaining part is power
suppressed. One notices that ${\rm Tr} \left [ U^{[\Box]}(0_T, y_T)
- U^{[\Box] \dag}(0_T,y_T) \right ]$ is in fact the dipole odderon
operator~\cite{Hatta:2005as}. The spin-dependent odderon has been
considered in this way in~\cite{Zhou:2013gsa} and in many studies of
elastic scattering, but without reference to TMDs, e.g.
\cite{Ryskin:1987ya,Buttimore:1998rj,Leader:1999ua}.

Next we use that the matrix element of the odderon operator only has
one possible $S_T$ dependence. It follows that for a transversely
polarized nucleon, the T-odd part of $\Gamma^{\mu \nu}$ can be
parameterized by only one leading-twist tensor structure~\cite{Zhou:2013gsa}:
\begin{equation}
\Gamma^{\mu \nu}_{\rm T-odd} (x, k_T; S_T) =  \frac{k_T^\mu k_T^\nu
N_c}{2 \pi^2 \alpha_s  x} \frac{\epsilon_{T}^{\alpha \beta}S_{T
\alpha}  k_{T \beta}}{M} O_{1T}^\perp(x,k_T^2),
 \end{equation}
where $O_{1T}^\perp(x,k_T^2)$ is identified as a spin dependent
odderon in~\cite{Zhou:2013gsa}. This leads us to identify:
\begin{eqnarray}
&& \!\!\!\!\!\!\!\!\!\!\! \frac{k_T^\mu k_T^\nu N_c}{2 \pi^2
\alpha_s  x} \frac{\epsilon_{T}^{\alpha \beta}S_{T \alpha}  k_{T
\beta}}{M} O_{1T}^\perp(x,k_T^2) = -\delta_T^{\mu \nu}
\frac{\epsilon_{T \alpha \beta} k_T^\alpha S_T^\beta}{M}
f_{1T}^{\perp g}
\nonumber \\
&& -\frac{ \tilde k_T^{ \{\mu } S_T^{\nu \} } +\tilde S_T^{ \{\mu }
k_T^{\nu \} }}{2 M}h_{1T}^{ g} + \frac{ \tilde k_T^{ \{\mu }
k_T^{\nu \} } }{ k_T^2 } \frac{k_T \cdot S_T}{ M}
h_{1T}^{\perp g} \,.
 \end{eqnarray}
In other words, the dipole type T-odd gluon TMDs satisfy
\begin{eqnarray}
xf_{1T}^{\perp g}=xh_{1T}^{ g}=x h_{1T}^{\perp g}=
\frac{-k_T^2 N_c}{4 \pi^2 \alpha_s } O_{1T}^\perp(x,k_T^2) \,,
 \end{eqnarray}
which is the main result of this Letter. We have now obtained a
consistent picture at small $x$ involving only one independent TMD,
determined by the expectation value of the spin-dependent odderon.
 We conclude that this one universal function, determined by
the imaginary part of a closed Wilson loop, should govern the single
transverse spin asymmetries in $p^\uparrow p $ and $p^\uparrow A$
scattering at RHIC in the
small-$x$ regime. This description differs from SSA involving the spin-{\it
in}dependent odderon \cite{Kovchegov:2012ga,Kovchegov:2013cva}.

The spin-dependent odderon has been considered in
\cite{Zhou:2013gsa} in the context of the McLerran-Venugopalan (MV)
model \cite{McLerran:1993ni}.  The MV model describes the small-$x$
distribution of gluons in the proton or nucleus as generated by a
Gaussian distribution of color sources.  The spin-dependent odderon
can be obtained \cite{Zhou:2013gsa} by including terms that are
cubic in the color sources~\cite{Hatta:2005as,Jeon:2005cf}. In
\cite{Zhou:2013gsa} it is observed that in the extended MV model
$O_{1T}^\perp(x,k_T^2)$ exhibits a node in $k_T$. Furthermore, our
tail calculations suggest that $k_T^2 O_{1T}^\perp(x,k_T^2)$ should
match onto a $1/k_T^4$ behavior at large $k_T$. About the $x$
dependence, it was noted in \cite{Zhou:2013gsa} that the evolution
of the odderon with increasing
energy~\cite{Bartels:1999yt,Bartels:1980pe} suggests that the
T-odd dipole gluon TMDs should fall off moderately with decreasing
$x$ as $x^{0.3}$ w.r.t.\ the unpolarized dipole gluon TMD (because the odderon has zero intercept, see also \cite{Kovchegov:2003dm}).
It should be feasible to test these expectations experimentally.
For this purpose one could study SSA in a number of processes that in the small-$x$ regime probe the dipole distributions, see e.g.\ \cite{Dominguez:2011wm}.
In $p^\uparrow A$ collisions these are backward hadron production~\footnote{As the odderon is $C$-parity odd,
for $gg$-dominated scattering one should select final states that are not $C$-even, i.e.\ $h^\pm \, X$
as opposed to $\pi^0 \ X$ or ${\rm jet}\, X$.}, $\gamma^*$ production, and
$\gamma^* \, {\rm jet}$ production in the back-to-back correlation limit.
BRAHMS data on SSA in backward charged hadron production
still allow for 10\% level asymmetries \cite{Arsene:2008aa}.

We end this Letter with some comments on the relation between the SSA
and parton orbital angular momentum. It is known that
the large-$x$ quark distribution in the transverse plane inside a
transversely polarized proton is distorted~\cite{Burkardt:2003uw}.
This distortion is related to the nonzero orbital angular momentum
of the quarks. As a consequence of this distortion, there will be a
left-right asymmetric distribution of color sources, resulting in an
asymmetric distribution of gluons at small $x$. This explains the
necessity of cubic source terms and the appearance of an odderon
contribution in a transversely polarized proton. In
\cite{Zhou:2013gsa} this idea was exploited to find a relation
between the spin-dependent odderon size and the anomalous magnetic
moments of the up and down quarks in the proton. Given these
insights, it is now also natural to expect that the three dipole
type T-odd gluon TMDs for general $x$ reflect features of the
distortion in the transverse plane of the gluon distribution in a
transversely polarized proton, and as such are an indirect
reflection of the presence of gluon orbital angular momentum.

In summary, we have calculated three leading power T-odd gluon TMDs
inside a transversely polarized hadron at large $k_T$ and in the
saturation regime. It has been found that the dipole type T-odd
gluon TMDs rise rapidly with decreasing $x$, whereas the WW type
ones are suppressed at small $x$. In deriving the latter, momentum
conservation was seen to play an essential role. This aspect remains
to be understood. The three dipole type T-odd gluon TMDs become
equal in the small-$x$ limit and are determined by the
spin-dependent odderon, which is given by the expectation value of a
single Wilson loop.
This leads to a surprisingly simple picture of transversely polarized hadrons at small $x$.\\

\noindent {\bf Acknowledgments:} J. Zhou thanks V. Braun for helpful discussions.
M.G.E. is supported by the ``Stichting voor Fundamenteel Onderzoek
der Materie'' (FOM), which is financially supported by the
``Nederlandse Organisatie voor Wetenschappelijk Onderzoek'' (NWO).
We acknowledge financial support from the European Community under
the FP7 "Ideas" program QWORK (contract 320389).

\end {document}